\documentclass[conference,a4paper]{IEEEtran}

\usepackage[T1]{fontenc}% optional T1 font encoding
\usepackage{graphicx}
\usepackage{titlesec}
\usepackage{setspace}

\titlespacing\section{0pt}{4pt plus 4pt minus 2pt}{4pt plus 2pt minus 2pt}
\titlespacing\subsection{0pt}{4pt plus 4pt minus 2pt}{4pt plus 2pt minus 2pt}
\titlespacing\subsubsection{0pt}{4pt plus 4pt minus 2pt}{4pt plus 2pt minus 2pt}
\usepackage{amsthm,amsmath,amssymb,bm}
\usepackage{mathrsfs}
\usepackage{amsfonts}

\usepackage[caption=false]{subfig}
\usepackage{algorithm}
\usepackage[noend]{algpseudocode}
\usepackage{graphics}
\usepackage{epsfig}
\usepackage{cite}
\usepackage{color}
\usepackage{booktabs}
\usepackage{verbatim}
\usepackage{stfloats}
\usepackage{cases}
\usepackage{multirow}
\usepackage{enumitem}
\usepackage{ragged2e}

\newcounter{MYtempeqncnt}

\newcommand{\transpose}{\mathsf{T}}
\newcommand{\hermconj}{\mathsf{H}}
\newcommand{\trace}{\mathtt{tr}}
\newcommand{\vect}{\mathrm{vec}}
\begin{document}
	
\title{Fluid Antenna-Assisted MU-MIMO Systems with Decentralized Baseband Processing}

\author{\IEEEauthorblockN{Tianyi~Liao, Wei~Guo, Hengtao~He, Shenghui~Song, \\ Jun~Zhang,~\textit{Fellow,~IEEE}, and Khaled~B.~Letaief,~\textit{Fellow,~IEEE}}
\IEEEauthorblockA{Dept. of ECE, The Hong Kong University of Science and Technology, Kowloon, Hong Kong\\ Emails: ty.liao@connect.ust.hk, \{eeweiguo, eehthe, eeshsong, eejzhang, eekhaled\}@ust.hk}}

\maketitle
% \IEEEaftertitletext{\vspace{1\baselineskip}}

\begin{abstract}
The fluid antenna system (FAS) has emerged as a disruptive technology, offering unprecedented degrees of freedom (DoF) for wireless communication systems. However, optimizing fluid antenna (FA) positions entails significant computational costs, especially when the number of FAs is large. To address this challenge, we introduce a decentralized baseband processing (DBP) architecture to FAS, which partitions the FA array into clusters and enables parallel processing. Based on the DBP architecture, we formulate a weighted sum rate (WSR) maximization problem through joint beamforming and FA position design for FA-assisted multiuser multiple-input multiple-output (MU-MIMO) systems. To solve the WSR maximization problem, we propose a novel decentralized block coordinate ascent (BCA)-based algorithm that leverages matrix fractional programming (FP) and majorization-minimization (MM) methods. The proposed decentralized algorithm achieves low computational, communication, and storage costs, thus unleashing the potential of the DBP architecture. Simulation results show that our proposed algorithm under the DBP architecture reduces computational time by over \(70\%\) compared to centralized architectures with negligible WSR performance loss.
\end{abstract} 

\begin{IEEEkeywords}
Fluid antenna system (FAS), decentralized baseband processing (DBP), MU-MIMO.
\end{IEEEkeywords}

\IEEEpeerreviewmaketitle

\setstretch{0.93}
\section{Introduction}\label{sec:intro}
The sixth-generation (6G) wireless communication systems aim to achieve terabit-per-second data rates, higher energy efficiency, and sub-millisecond latency~\cite{letaiefRoadmap6GAI2019}. To achieve this goal, massive multiple-input multiple-output (MIMO) and multiuser MIMO (MU-MIMO) will play pivotal roles. 
% The major advantage of MIMO systems is their ability to leverage spatial degrees of freedom (DoF), which can improve the system performance by exploiting spatial diversity and multiplexing gains~\cite{zhengDiversityMultiplexingFundamental2003}. 
However, conventional MIMO systems assume fixed-position antenna (FPA) configurations, which lack the flexibility to dynamically adapt to changing propagation environments and thus cannot fully exploit the spatial degrees of freedom (DoF).

To address these challenges, the fluid antenna system (FAS) was proposed in~\cite{wong2020fluid} as a disruptive technology. 
%FAS utilizes fluid antennas (FAs) to achieve flexible control over the positions, gains, radiation patterns, and other characteristics of the antennas~\cite{wongFluidAntennaSystems2021}. 
Among various categories of FAS, position optimization has proven to be an effective approach for fully harnessing the spatial DoF. The first FA position optimization algorithm was introduced in~\cite{maMIMOCapacityCharacterization2024}, where the authors considered a rate maximization problem in a point-to-point MIMO system and solved it with a successive concave approximation (SCA)-based algorithm. Several subsequent works demonstrated the benefits of position optimization in FAS from various perspectives, including reductions in transmit power~\cite{zhuMovableAntennaEnhancedMultiuser2024}, guarantees of minimum transmission rates~\cite{fengWeightedSumRateMaximization2024}, and enhancements in physical layer security~\cite{tangSecureMIMOCommunication2024}. %Recently, the paper~\cite{fengWeightedSumRateMaximization2024} addressed a weighted sum rate (WSR) maximization problem with jointly optimizing FA positions and beamforming matrices in a multiuser multiple-input single-output system. The joint optimization significantly improved system capacity of FAS compared to conventional MIMO systems with fixed-position antennas (FPAs).

However, the complexity of solving the joint optimization problem scales cubically with the number of transmit FAs~\cite{fengWeightedSumRateMaximization2024}, leading to significant \textit{computational costs}. Since the high computational cost stems from the large number of FAs, a practical solution is to decompose the optimization problem into smaller subproblems. The decentralized baseband processing (DBP) architecture~\cite{li2017decentralized} achieves this goal by partitioning large transmit antenna arrays into clusters, with each cluster managed by a decentralized unit (DU). By enabling parallel computation, the DBP architecture substantially reduces the \textit{computational costs} of baseband processing algorithms while ensuring feasible \textit{communication} and \textit{storage costs} for each hardware unit. As a result, the DBP architecture has been extensively utilized for beamforming~\cite{zhaoCommunicationEfficientDecentralizedLinear2023}, channel estimation~\cite{xu2023low}, and signal detection~\cite{zhuang2025decentralized} in conventional MIMO systems.

Despite the advantages of the DBP architecture in conventional MIMO systems, its efficient implementation in FA-assisted MU-MIMO systems remains unexplored. To bridge this gap, we propose incorporating the DBP architecture into the FA-assisted MU-MIMO system and consider the weighted sum rate (WSR) maximization problem via joint beamforming and FA position optimization. However, the benefits of the DBP architecture cannot be achieved without efficient decentralized optimization algorithms. 
% \textcolor{blue}{Different from conventional MIMO systems, joint optimization in FA-assisted MU-MIMO systems relies heavily on global knowledge of channel matrices, beamforming matrices, and transmit FA positions that is not directly accessible under the DBP architecture. To address this challenge, we design efficient communication schemes between the centralized unit (CU) and the decentralized units (DUs), enabling access to the necessary global information while ensuring low \textit{communication} and \textit{storage costs}. Based on the communication scheme}, 
Therefore, we develop a decentralized block coordinate ascent (BCA)-based framework to solve the WSR maximization problem. Specifically, we first decouple the optimization of beamforming matrices and FA positions by leveraging fractional programming (FP) techniques. To enable the optimization of beamforming matrices under the DBP architecture, we adopt the Nesterov's extrapolation and matrix non-homogeneous transform~\cite{zhangDiscerningEnhancingWeighted2023}, which efficiently eliminates matrix inversions and enables each DU to optimize its beamforming matrix independently. Different from the decentralized algorithms for conventional MIMO systems, we introduce a novel majorization-minimization (MM) algorithm for distributed FA position optimization where each DU can simultaneously optimize the transmit FAs it manages. The proposed decentralized algorithm achieves low \textit{computational cost} by enabling all DUs to compute in parallel. In addition, the decentralized BCA-based framework ensures low \textit{communication} and \textit{storage costs}.

%Simulation results demonstrate that the DBP architecture reduces the computational cost of the BCA-based algorithm by over $70\%$, while maintaining a WSR performance loss of less than $3\%$.

The remainder of this paper is organized as follows. Section~\ref{sec:system} presents the system model of the FA-assisted MU-MIMO networks with the DBP architecture and formulates the WSR maximization problem. Section~\ref{sec:bca_sol} reformulates the problem using FP techniques and solves it using the decentralized BCA-based algorithm. Simulation results are provided in Section~\ref{sec:sim}, and conclusions are drawn in Section~\ref{sec:conclusion}.

% \textit{Notation:} For a complex number $a$, its amplitude and phase are given by $\lvert a \rvert$ and $\angle a$, respectively. The $\ell_2$ norm of a vector $\mathbf{a}$ is $\lVert \mathbf{a} \rVert_2$. $[\mathbf{A}]_m$, $[\mathbf{A}]_{mn}$, $\mathbf{A}^\transpose$, $\mathbf{A}^\hermconj$, $\det(\mathbf{A})$, $\trace(\mathbf{A})$, $\vect(\mathbf{A})$,  $\lVert \mathbf{A} \rVert_\infty$, and $\lVert \mathbf{A} \rVert_{\rm F}$ denote the $m$-th row, the $(m, n)$-th element, transpose, conjugate transpose, determinant, trace, vectorization, the infinity norm, and the Frobenius norm of matrix $\mathbf{A}$, respectively. $\mathbf{A} \succeq \mathbf{0}$ and $\mathbf{A} \succ \mathbf{0}$ indicate that $\mathbf{A}$ is positive semi-definite and positive definite, respectively. $\mathbb{C}^{M \times N}$, $\mathbb{R}^{M \times N}$, and $\mathbb{R}_{+}^{M \times N}$ denote the sets of $M \times N$ complex, real, and non-negative real matrices, respectively. The circularly symmetric complex Gaussian (CSCG) distribution with zero mean and covariance $\sigma^2 \mathbf{I}$ is represented as $\mathcal{CN}(\mathbf{0}, \sigma^2 \mathbf{I})$, and the uniform distribution over $[a, b]$ is denoted by $\mathcal{U}[a, b]$.

\section{System Model and Problem Formulation}\label{sec:system}
\begin{figure*}[tb]
    \centering
    \includegraphics[width=\textwidth]{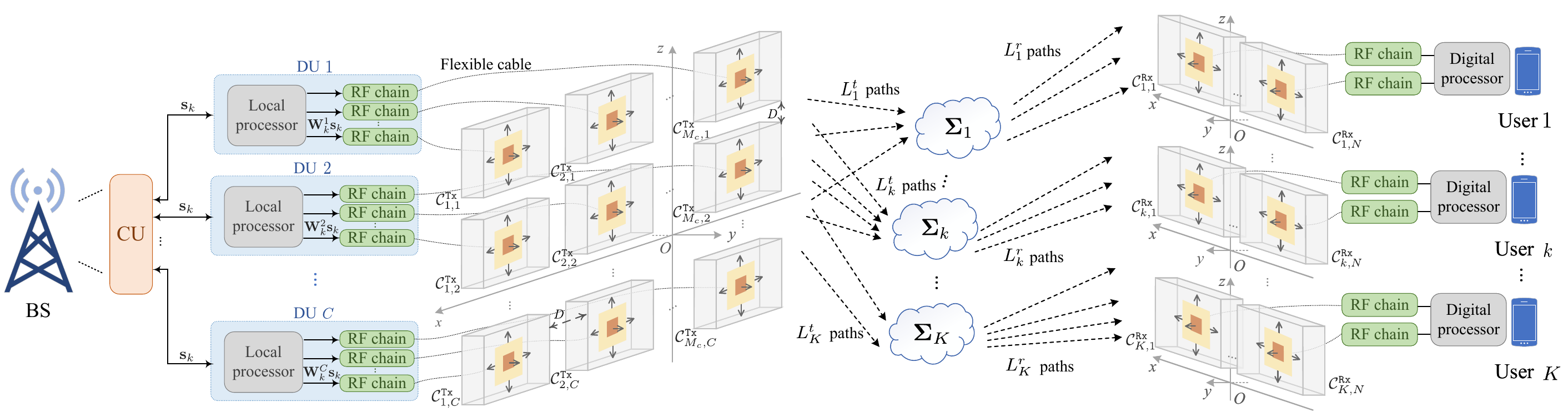}
    \caption{The system model of the FA-assisted downlink MU-MIMO with the DBP architecture at the BS.}
    \vspace{-1\baselineskip}
    \label{fig:system}
\end{figure*}
As shown in Fig.~\ref{fig:system}, we consider a downlink MU-MIMO system where a BS with $M$ FAs serves $K$ users, each equipped with $N$ FAs. The BS employs a DBP architecture consisting of $C$ clusters, and all of them are connected to a centralized unit (CU). Each cluster includes $M_c = M / C$ transmit FAs and a decentralized unit (DU) equipped with a local processor and RF chains~\cite{zhaoCommunicationEfficientDecentralizedLinear2023}. Let $\mathbf{s}_k \in \mathbb{C}^{d}$ denote the data stream intended for user $k$, where $d \leq \min\{M, N\}$ is the number of parallel data streams. The beamforming is performed at the DUs, and the $c$-th DU uses the beamforming matrix $\mathbf{W}_k^c \in \mathbb{C}^{M_c \times d}$ for transmitting $\mathbf{s}_k$ from the BS to user $k$. %To initiate transmission, the CU first broadcasts $\mathbf{s}_k$ to all DUs, after which each DU applies beamforming and transmits the precoded signal. The signal transmitted from the $c$-th DU to user $k$ is given by $\mathbf{x}_k^c = \mathbf{W}_k^c \mathbf{s}_k$.

To describe the positions of the transmit FAs at the BS and receive FAs at the users, we establish a three-dimensional (3D) Cartesian coordinate system. Specifically, let $\mathbf{t}_m^c =[x_{m, c}^\mathtt{Tx}, y_{m, c}^\mathtt{Tx}, z_{m, c}^\mathtt{Tx}]^\transpose$ denote the position of $m$-th transmit FA at the $c$-th DU of the BS and let $\mathbf{r}_{kn} =[x_{kn}^\mathtt{Rx}, y_{kn}^\mathtt{Rx}, z_{kn}^\mathtt{Rx}]^\transpose$ denote the position of the $n$-th receive FA at user $k$, where $\mathcal{C}_{m, c}^\mathtt{Tx}$ and $\mathcal{C}_{kn}^\mathtt{Rx}$ are the given 3D movable regions of transmit and receive FAs and are assumed to be cuboid~\cite{zhuMovableAntennaEnhancedMultiuser2024}. Moreover, to mitigate the coupling effect between different FAs, we assume that any pair of transmit and receive movable regions are non-overlapping and separated by a distance of at least $D$.~\footnote{This assumption simplifies the optimization problem, enhances practical feasibility for engineering implementation, and results in negligible performance loss~\cite{liao2025joint}.} 
% In this paper, we consider narrow-band slow fading channels, where the transmit and receive FAs remain static or move slowly within the movable region during each quasi-static fading block. Moreover, the time required for FA movement is assumed to be much shorter than the coherence time~\cite{maMIMOCapacityCharacterization2024}. 
Therefore, the received signal $\mathbf{y}_k \in \mathbb{C}^N$ at user $k$ is given by
\begin{small}
\begin{equation}\label{eq:signal_model}
    \mathbf{y}_k = \sum_{c=1}^C{\mathbf{H}_k^c(\mathbf{T}^c, \mathbf{R}_k) \mathbf{W}_k^c} \mathbf{s}_k + \sum_{j=1, j\neq k}^K{\mathbf{H}_k^c(\mathbf{T}^c, \mathbf{R}_k)\mathbf{W}_j^c} \mathbf{s}_j + \mathbf{n}_k,
\end{equation}
\end{small}%
where $\mathbf{T}^c = [\mathbf{t}_1^c, \dots, \mathbf{t}_{M_c}^c]^\transpose$ represents the positions of the transmit FAs at the $c$-th DU and is stored locally at the $c$-th DU, and $\mathbf{R}_k = [\mathbf{r}_{k1}, \dots, \mathbf{r}_{kN}]^\transpose$ denotes the positions of the receive FAs at user $k$ and is stored at the CU. The channel matrix between the FAs at the $c$-th DU of the BS and user $k$ is given by $\mathbf{H}_k^c(\mathbf{T}^c, \mathbf{R}_k) \in \mathbb{C}^{N\times M_c}$, which depends on both $\mathbf{T}^c$ and $\mathbf{R}_k$. The term $\mathbf{n}_k \in \mathbb{C}^{N}$ represents additive white Gaussian noise (AWGN) following the distribution $\mathcal{CN}(\mathbf{0}, \sigma_k^2 \mathbf{I})$.

\subsection{Channel Model}\label{subsec:channel}
In this paper, we apply the field response model in the far-field~\cite{maMIMOCapacityCharacterization2024}. Let $L_k^\mathtt{Tx}$ and $L_k^\mathtt{Rx}$ denote the numbers of transmit and receive channel paths between the BS and user $k$, respectively. The steering vectors corresponding to the $q$-th transmit and receive paths are given by
\begin{align}\label{eq:steering_vector}
	\mathbf{g}_{kq}^\mathtt{Tx}&=\left[\cos\theta_{kq}^\mathtt{Tx}\cos\phi_{kq}^\mathtt{Tx}, \cos\theta_{kq}^\mathtt{Tx}\sin\phi_{kq}^\mathtt{Tx}, \sin\theta_{kq}^\mathtt{Tx}\right]^{\transpose},\\
	\mathbf{f}_{kq}^\mathtt{Rx}&=\left[\cos\theta_{kq}^\mathtt{Rx}\cos\phi_{kq}^\mathtt{Rx},\cos\theta_{kq}^\mathtt{Rx}\sin\phi_{kq}^\mathtt{Rx},\sin\theta_{kq}^\mathtt{Rx}\right]^{\transpose},
\end{align}
where $\theta_{kq}^\mathtt{Tx}$ and $\phi_{kq}^\mathtt{Tx}$ (and $\theta_{kq}^\mathtt{Rx}$ and $\phi_{kq}^\mathtt{Rx}$) are the elevation and azimuth angle of arrivals (and angle of departures) of the $q$-th path between the BS and user $k$. For the $q$-th transmit (and receive) channel path from the BS to user $k$, the distance difference between the path originating from the $m$-th BS antenna position at the $c$-th DU $\mathbf{t}_m^c$ (and $n$-th user antenna position $\mathbf{r}_{kn}$) and that from the origin of the BS (and user $k$) coordinate system are given by
\begin{equation}\label{eq:distance_diff}
    \rho_{kq}^\mathtt{Tx}(\mathbf{t}_m^c) \triangleq \left(\mathbf{g}_{kq}^\mathtt{Tx}\right)^\transpose \mathbf{t}_m^c, \qquad
    \rho_{kq}^\mathtt{Rx}(\mathbf{r}_{kn}) \triangleq \left(\mathbf{f}_{kq}^\mathtt{Rx}\right)^\transpose \mathbf{r}_{kn},
\end{equation}
% Similarly, for the $j$-th receive channel path from the BS to user $k$, the distance difference between the path originating from the $n$-th receive antenna position $\mathbf{r}_{kn}$ and that from the origin of user $k$'s coordinate system $\mathbf{o}_k$ is given by
% \begin{equation}\label{eq:distance_diff_user_k}
%     \rho_{kj}^\mathtt{Rx}(\mathbf{r}_{kn})\triangleq\left(\mathbf{f}_{kj}^\mathtt{Rx}\right)^\transpose \mathbf{r}_{kn}.
% \end{equation} 
respectively. The transmit and receive field-response vectors (FRVs) between the BS and user $k$ are given by~\cite{zhuMovableAntennaEnhancedMultiuser2024}
\begin{align}
	\mathbf{g}^c_k(\mathbf{t}_m^c)&\triangleq\big[\mathrm{e}^{\jmath\frac{2\pi}{\lambda}\rho_{k1}^\mathtt{Tx}(\mathbf{t}_m)},\dots, \mathrm{e}^{\jmath\frac{2\pi}{\lambda}\rho_{k,L_k^\mathtt{Tx}}^\mathtt{Tx}(\mathbf{t}_m^c)}\big]^{\transpose},\label{eq:tx_FRV}\\
	\mathbf{f}_k(\mathbf{r}_{kn})&\triangleq\big[\mathrm{e}^{\jmath\frac{2\pi}{\lambda}\rho_{k1}^\mathtt{Rx}(\mathbf{r}_{kn})},\cdots, \mathrm{e}^{\jmath\frac{2\pi}{\lambda}\rho_{k,L_k^\mathtt{Rx}}^\mathtt{Rx}(\mathbf{r}_{kn})}\big]^{\transpose},\label{eq:rx_FRV}
\end{align} 
respectively, where $\lambda$ denotes the carrier wavelength. By defining the path-response matrix (PRM) $\mathbf{\Sigma}_k\in\mathbb{C}^{L_k^r\times L_k^t}$ as the response between each pair of transmit and receive channel paths from the BS to user $k$, the channel matrix $\mathbf{H}_k^c\left(\mathbf{T}^c,\mathbf{R}_k\right)$ is given by
\begin{equation}\label{eq:channel_k}
	\mathbf{H}_k^c\left(\mathbf{T}^c, \mathbf{R}_k\right) = \mathbf{F}_k^{\hermconj}\left(\mathbf{\mathbf{R}_k}\right)\mathbf{\Sigma}_k\mathbf{G}_k^c\left(\mathbf{T}^c\right),
\end{equation}
where $\mathbf{F}_k\left(\mathbf{R}_k\right)=\left[\mathbf{f}_k(\mathbf{r}_{k1}),\dots,\mathbf{f}_k(\mathbf{r}_{kn})\right]$ and $\mathbf{G}_k^c\left(\mathbf{T}^c\right)=\left[\mathbf{g}^c_k(\mathbf{t}_1^c),\cdots,\mathbf{g}^c_k(\mathbf{t}_M^c)\right]$ denote the field response matrices (FRMs) of all the receive FAs at user $k$ and those at the BS, respectively.

\subsection{Problem Formulation}\label{subsec:problem}
A fundamental problem in MU-MIMO downlink transmission is WSR maximization. The WSR is defined as  
\begin{equation}\label{eq:def_wsr}
	R=\sum\nolimits_{k=1}^{K}\alpha_k R_k,
\end{equation}
where the weight $\alpha_k$ denotes the priority of user $k$. $R_k$ is the achievable rate of user $k$, given by
\begin{small}
\begin{equation}\label{eq:achievable_rate}
    R_k = \log\det\bigg(\mathbf{I} + \Big(\sum_{c=1}^C{\mathbf{H}_k^c}\mathbf{W}_k^c\Big)^\hermconj \mathbf{M}_k^{-1} \Big(\sum_{c=1}^C{\mathbf{H}_k^c}\mathbf{W}_k^c\Big)\bigg),
\end{equation}
\end{small}%
according to the signal model with the DBP architecture in~\eqref{eq:signal_model}. The interference-plus-noise matrix $\mathbf{M}_k$ is defined as
\begin{small}
\begin{equation}\label{eq:snr_matrix}
	\mathbf{M}_k = \sum_{j=1, j\neq k}^{K}\Big(\sum_{c=1}^C \mathbf{H}_k^c \mathbf{W}_j^c\Big)\Big(\sum_{c=1}^C\mathbf{H}_k^c \mathbf{W}_j^c\Big)^\hermconj + \sigma_k^2\mathbf{I}.
\end{equation}
\end{small}%
Let $\underline{\mathbf{W}}=\{\mathbf{W}_k^c,\forall k,c\}$, $\underline{\mathbf{T}}=\{\mathbf{T}^c,\forall c\}$ and $\underline{\mathbf{R}}=\{\mathbf{R}_k,\forall k\}$ denote the set of beamforming matrices, transmit, and receive FA positions, respectively. Then, we can formulate the optimization problem as
\begin{subequations}\label{opt:problem}
	\begin{align} 			   
		\underset{\underline{\mathbf{W}}, \underline{\mathbf{T}}, \underline{\mathbf{R}}}{\max}\ \ &R\label{opt:obj}\\
		{\text{s. t.}}\ \ \ &\sum\nolimits_{k=1}^{K}\sum\nolimits_{c=1}^C \trace\left(\mathbf{W}_k^c\left(\mathbf{W}_k^c\right)^\hermconj\right) \leq P_{\max},\label{opt:power_constraint}\\
		&\mathbf{t}_{m, c}\in\mathcal{C}_{m, c}^{\mathtt{Tx}},\ \forall m,\label{opt:tx_antenna_constraint}\\ 
		&\mathbf{r}_{kn}\in\mathcal{C}_{kn}^{\mathtt{Rx}},\ \forall{kn},\label{opt:rx_antenna_constraint}
	\end{align}
\end{subequations} 
where $P_{\max}$ denotes the total transmit power budget at the BS. The problem~\eqref{opt:problem} is difficult to solve because the objective function~\eqref{opt:obj} is highly non-concave and the optimization variables are highly coupled. 

\section{Decentralized Block Coordinate Ascent (BCA)-Based Algorithm}\label{sec:bca_sol}
In this section, we propose a decentralized BCA-based algorithm to solve problem~\eqref{opt:problem}. First, we employ the FP method to decouple the variables in problem~\eqref{opt:problem}. Then, we introduce a decentralized inverse-free algorithm to optimize the beamforming matrices. Furthermore, a decentralized MM algorithm is proposed to handle the non-concave optimization of FA positions.

\subsection{Decentralized Matrix Multiplication under DBP}
\begin{figure*}[hb]
	\normalsize
    \vspace{-1\baselineskip}
    \hrulefill
    \setcounter{MYtempeqncnt}{\value{equation}}
    \setcounter{equation}{12}
    \begin{small}
	\begin{multline}\label{eq:quadratic_trans}
        f_{\rm Quad}\left(\underline{\mathbf{W}}, \underline{\mathbf{T}}, \underline{\mathbf{R}}, \underline{\mathbf{\Gamma}}, \underline{\mathbf{\Phi}}\right)=
        \sum_{k=1}^{K}\left(\alpha_k\log\det\left(\mathbf{I}+\mathbf{\Gamma}_k\right)-\alpha_k\trace\left(\mathbf{\Gamma}_k\right) +\trace\left(\left(\mathbf{I}+\mathbf{\Gamma}_k\right)\left(\sqrt{\alpha_k}\left(\sum_{c=1}^C\mathbf{H}_k^c(\mathbf{T}^c, \mathbf{R}_k)\mathbf{W}_k^c\right)^\hermconj\mathbf{\Phi}_k \right.\right.\right.\\
        \left.\left.\left.+ \sqrt{\alpha_k}\mathbf{\Phi}_k^{\hermconj}\sum_{c=1}^C\mathbf{H}_k^c(\mathbf{T}^c, \mathbf{R}_k)\mathbf{W}_k^c - \mathbf{\Phi}_k^{\hermconj}\left(\sum_{j=1}^{K}\left(\sum_{c=1}^C\mathbf{H}_k^c(\mathbf{T}^c, \mathbf{R}_k)\mathbf{W}_k^c\right)\left(\sum_{c=1}^C\mathbf{H}_k^c(\mathbf{T}^c, \mathbf{R}_k)\mathbf{W}_k^c\right)^\hermconj + \sigma_k^2\mathbf{I}\right)\mathbf{\Phi}_k\right)\right)\right).
	\end{multline}
    \end{small}%
    \setcounter{equation}{\value{MYtempeqncnt}}
\end{figure*}

Since the number of FAs at the BS $M$ is significantly larger than that at the user side $N$, it is essential to ensure that the computational complexity of all operations is independent of $M$ to maintain low \textit{computational costs}. For the same reason, the dimensions of matrices stored at the CU and DUs should also be independent of $M$, thereby ensuring low \textit{storage} and \textit{communication costs}.

\begin{figure}[h]
    \vspace{-0.4cm}
    \centering
    \includegraphics[width=0.35\textwidth]{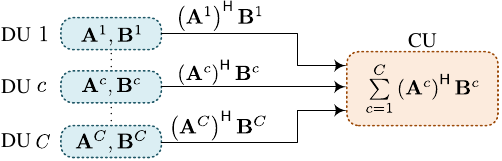}
    \caption{The decentralized calculation of $\sum_{c=1}^C (\mathbf{A}^c)^\hermconj \mathbf{B}^c$.}
    \vspace{-0.4cm}
    \label{fig:mul}
\end{figure}

To meet these requirements, efficient decentralized computation methods must be designed under the DBP architecture. Among them, decentralized matrix multiplication is frequently utilized in the proposed BCA-based algorithm, as discussed later. Given its pivotal role in our proposed algorithm, we first formalize this operation along with its implementation before presenting the algorithmic details. Suppose matrices $\mathbf{A}^c \in \mathbb{C}^{M_c \times a}$ and $\mathbf{B}^c \in \mathbb{C}^{M_c \times b}$ are stored at the $c$-th DU, where $a, b \in \mathbb{N}^+$. We define the function $\mathsf{Mul}(\cdot, \cdot)$ as
\begin{equation}
    \mathsf{Mul}(\mathbf{A}^c, \mathbf{B}^c) = \sum\nolimits_{c=1}^C (\mathbf{A}^c)^\hermconj \mathbf{B}^c,
\end{equation}
whose result has dimensions independent of $M$ and can be efficiently stored and transmitted between the CU and DUs. The decentralized computation of $\mathsf{Mul}(\mathbf{A}^c, \mathbf{B}^c)$ is illustrated in Fig.~\ref{fig:mul}. Specifically, each DU first computes $(\mathbf{A}^c)^\hermconj \mathbf{B}^c$ locally and then transmits the result to the CU, where the final sum is aggregated.

\subsection{Problem Reformulation}\label{subsec:problem_reform}
Next, we reformulate problem~\eqref{opt:problem} into a more tractable form. Since the objective function~\eqref{opt:obj} is a sum-of-functions-of-matrix-ratios, we apply the matrix FP framework developed in~\cite{shen2019optimization} to decouple different sets of variables. By utilizing matrix Lagrangian dual transform~\cite[Theorem~2]{shen2019optimization} and matrix quadratic transform~\cite[Theorem~1]{shen2019optimization} to the problem~\eqref{opt:problem}, we introduce two sets of auxiliary variables, $\underline{\mathbf{\Gamma}}=\{\mathbf{\Gamma}_k,\forall k\}$ and $\underline{\mathbf{\Phi}}=\{\mathbf{\Phi}_k,\forall k\}$, and reformulate the problem~\eqref{opt:problem} as
\begin{subequations}\label{opt:problem_quadratic}
	\begin{align} 			   
		\underset{\underline{\mathbf{W}}, \underline{\mathbf{T}}, \underline{\mathbf{R}}, \underline{\mathbf{\Gamma}}, \underline{\mathbf{\Phi}}}{\max}\ \ &f_{\rm Quad}\left(\underline{\mathbf{W}}, \underline{\mathbf{T}}, \underline{\mathbf{R}}, \underline{\mathbf{\Gamma}}, \underline{\mathbf{\Phi}}\right)\label{opt:obj_quadratic}\\
		{\text{s. t.}}\ \ \ \ \ &\eqref{opt:power_constraint}-\eqref{opt:rx_antenna_constraint},
	\end{align}
\end{subequations} 
where $f_{\rm Quad}\left(\underline{\mathbf{W}}, \underline{\mathbf{T}}, \underline{\mathbf{R}}, \underline{\mathbf{\Gamma}}, \underline{\mathbf{\Phi}}\right)$ is given by~\eqref{eq:quadratic_trans} at the bottom of the page. 

\addtocounter{equation}{1}

Then, we propose a decentralized BCA-based algorithm to solve the problem~\eqref{opt:problem_quadratic} by iteratively optimizing one set of variables while keeping others fixed until convergence.

\subsection{Decentralized Update of $\underline{\mathbf{\Gamma}}$ and $\underline{\mathbf{\Phi}}$}\label{subsec:auxiliary}
In this step, we aim to optimize the auxiliary variables $\underline{\mathbf{\Gamma}}$ and $\underline{\mathbf{\Phi}}$ with fixed $\underline{\mathbf{W}}$, $\underline{\mathbf{T}}$, and $\underline{\mathbf{R}}$. The values of $\underline{\mathbf{\Gamma}}$ and $\underline{\mathbf{\Phi}}$ are computed and stored at the CU. By setting the first-order derivative of~\eqref{eq:quadratic_trans} to zero with respect to (w.r.t.) $\mathbf{\Gamma}_k$ and $\mathbf{\Phi}_k$, we obtain the closed-form expressions for the optimal $\mathbf{\Gamma}_k$ and $\mathbf{\Phi}_k$, given by
\begin{align}
    \mathbf{\Gamma}_k &= \tilde{\mathbf{G}}_{kk}^\hermconj\mathbf{\Sigma}_k^\hermconj\overline{\mathbf{F}}_k \overline{\mathbf{M}}_k^{-1}\overline{\mathbf{F}}_k^\hermconj\mathbf{\Sigma}_k\tilde{\mathbf{G}}_{kk}, \label{eq:gamma}\\
    \mathbf{\Phi}_k &= \sqrt{\alpha_k}\left(\overline{\mathbf{M}}_k+\overline{\mathbf{F}}_k^\hermconj\mathbf{\Sigma}_k\tilde{\mathbf{G}}_{kk}\tilde{\mathbf{G}}_{kk}^\hermconj\mathbf{\Sigma}_k^\hermconj\overline{\mathbf{F}}_k\right)^{-1}\overline{\mathbf{F}}_k^\hermconj\mathbf{\Sigma}_k\tilde{\mathbf{G}}_{kk},\label{eq:phi}
\end{align}
where $\overline{\mathbf{T}}^c$, $\overline{\mathbf{W}}_j^c$, $\overline{\mathbf{F}}_k$, and $\overline{\mathbf{M}}_k$ denote the temporally optimized values obtained from the previous iteration, and $\tilde{\mathbf{G}}_{kj} \triangleq \mathsf{Mul}\Big(\big(\mathbf{G}^c_k(\overline{\mathbf{T}}^c)\big)^\hermconj, \overline{\mathbf{W}}_j^c\Big)$. After computing $\tilde{\mathbf{G}}_{kj}$ distributively, which is similar to the process in Fig.~\ref{fig:mul}, the remaining calculations of $\mathbf{\Gamma}_k$ and $\mathbf{\Phi}_k$ are performed directly at the CU since the dimensions of $\mathbf{\Gamma}_k$ and $\mathbf{\Phi}_k$ are independent of $M$.

\subsection{Decentralized Update Step of $\underline{\mathbf{W}}$}\label{subsec:W}
In this step, we aim to optimize the beamforming matrices $\underline{\mathbf{W}}$ with fixed $\underline{\mathbf{\Gamma}}$, $\underline{\mathbf{\Phi}}$, $\underline{\mathbf{T}}$, and $\underline{\mathbf{R}}$. 
Under the DBP architecture, the value of $\underline{\mathbf{W}}$ are computed at the DUs. However, optimizing beamforming matrices by directly using FP techniques requires matrix inversion operations~\cite{shen2019optimization}, which are incompatible with the DBP architecture since matrix inversions cannot be performed distributively at the DUs. To update all $\mathbf{W}_k^c$ locally at each DU, we employ Nesterov's extrapolation and the matrix non-homogeneous transform to efficiently eliminate the matrix inversion operation.

% \begin{figure*}[hb]
%     \normalsize
%     \hrulefill
%     \setcounter{MYtempeqncnt}{\value{equation}}
%     \setcounter{equation}{25}
%     \begin{equation}\label{eq:opt_tx_position_delta}
%         \begin{aligned}
%             \delta^{\mathtt{Tx}} = \underset{1\leq m\leq M}{\max}\frac{24\pi^2}{\lambda^2}&\sum\nolimits_{k=1}^K L_k^t\left(\left(\sum\nolimits_{t=1}^K\Big\lVert[\overline{\mathbf{W}}_t^c]_m\Big\rVert_2 \sum\nolimits_{j=1}^M \Big\lVert[\overline{\mathbf{W}}_t]_j\Big\rVert_2 + \sum\nolimits_{t=1}^K\sum\nolimits_{s=1}^K[\overline{\mathbf{W}}_t^c]_m\tilde{\mathbf{W}}_{ts}[\overline{\mathbf{W}}_s^c]_m^\hermconj\right)\right.\\
%             &\left.\times \lVert\mathbf{\Sigma}_k^\hermconj \overline{\mathbf{F}}_k \overline{\mathbf{\Phi}}_k \left(\mathbf{I} + \overline{\mathbf{\Gamma}}_k\right) \overline{\mathbf{\Phi}}_k^\hermconj \overline{\mathbf{F}}_k^\hermconj \mathbf{\Sigma}_k\rVert_2 + \sqrt{\frac{\alpha_k}{L_k^t}} \Big\lVert[\overline{\mathbf{W}}_k^c]_m\left(\mathbf{I}+\overline{\mathbf{\Gamma}}_k\right)\overline{\mathbf{\Phi}}_k^\hermconj\overline{\mathbf{F}}_k^\hermconj\mathbf{\Sigma}_k^\hermconj\Big\rVert_2 \right).
%         \end{aligned}    
%     \end{equation}
%     \setcounter{equation}{\value{MYtempeqncnt}}
% \end{figure*}

\begin{figure*}[hb]
    \normalsize
    \vspace{-1\baselineskip}
    \hrulefill
    \setcounter{MYtempeqncnt}{\value{equation}}
    \setcounter{equation}{25}
    \begin{equation}\label{eq:opt_tx_position_delta}
        \begin{aligned}
            \delta^{\mathtt{Tx}}_{m, c} = \frac{24\pi^2}{\lambda^2}&\sum\nolimits_{k=1}^K L_k^t\left(\left(\sum\nolimits_{t=1}^K\Big\lVert[\overline{\mathbf{W}}_t^c]_m\Big\rVert_2 \sum\nolimits_{j=1}^M \Big\lVert[\overline{\mathbf{W}}_t^c]_j\Big\rVert_2 + \sum\nolimits_{t=1}^K\sum\nolimits_{s=1}^K[\overline{\mathbf{W}}_t^c]_m\tilde{\mathbf{W}}_{ts}[\overline{\mathbf{W}}_s^c]_m^\hermconj\right)\right.\\
            &\left.\times \lVert\mathbf{\Sigma}_k^\hermconj \overline{\mathbf{F}}_k \overline{\mathbf{\Phi}}_k \left(\mathbf{I} + \overline{\mathbf{\Gamma}}_k\right) \overline{\mathbf{\Phi}}_k^\hermconj \overline{\mathbf{F}}_k^\hermconj \mathbf{\Sigma}_k\rVert_2 + \sqrt{\frac{\alpha_k}{L_k^t}} \Big\lVert[\overline{\mathbf{W}}_k^c]_m\left(\mathbf{I}+\overline{\mathbf{\Gamma}}_k\right)\overline{\mathbf{\Phi}}_k^\hermconj\overline{\mathbf{F}}_k^\hermconj\mathbf{\Sigma}_k^\hermconj\Big\rVert_2 \right).
        \end{aligned}    
    \end{equation}
    \setcounter{equation}{\value{MYtempeqncnt}}
\end{figure*}

To obtain the closed-form expression of $\mathbf{W}_k^c$ without the need for matrix inversion operations, we first apply Nesterov's extrapolation to predict $\mathbf{W}_k^c$ in the next iteration using the two previous ones, $\overline{\mathbf{W}}_k^c$ and $\overline{\overline{\mathbf{W}}}_k^c$. The extrapolated value $\mathbf{\Upsilon}_k^c$ is defined as
\begin{equation}\label{eq:Upsilon}
    \mathbf{\Upsilon}_k^c \triangleq \overline{\mathbf{W}}_k^c + \nu_i\left(\overline{\mathbf{W}}_k^c - \overline{\overline{\mathbf{W}}}_k^c\right),
\end{equation}
where $\nu_i = \max\left\{(i-2)/(i+1), 0\right\}$ represents the extrapolation step size in the $i$-th BCA iteration. 
% For simplicity, we define
% \begin{equation}\label{eq:HxUpsilon}
%     \tilde{\mathbf{\Upsilon}}_{jk} \triangleq \sum\nolimits_{c=1}^C \overline{\mathbf{H}}_j^c \mathbf{\Upsilon}_k^c.
% \end{equation}
Next, by applying the matrix non-homogeneous transform to~\eqref{eq:quadratic_trans}, we obtain the beamforming matrix at the $c$-th DU without the power constraint, expressed as~\cite{liao2025joint}
\begin{equation}\label{eq:extr_q}
    \begin{aligned}
        \mathbf{Q}_k^c =& \eta^{-1}\sqrt{\alpha_k}(\overline{\mathbf{H}}_k^c)^\hermconj\overline{\mathbf{\Phi}}_k\left(\mathbf{I} + \overline{\mathbf{\Gamma}}_k\right)\\ 
        &- \eta^{-1}\sum\nolimits_{j=1}^{K}(\overline{\mathbf{H}}_j^c)^\hermconj\overline{\mathbf{\Phi}}_j\left(\mathbf{I} + \overline{\mathbf{\Gamma}}_j\right)\overline{\mathbf{\Phi}}_j^\hermconj\tilde{\mathbf{\Upsilon}}_{jk} + \mathbf{\Upsilon}_k^c,
    \end{aligned}
\end{equation}
where $\tilde{\mathbf{\Upsilon}}_{jk} \triangleq \mathsf{Mul}\left((\overline{\mathbf{H}}_j^c)^\hermconj, \mathbf{\Upsilon}_k^c\right)$, $\overline{\mathbf{H}}_k^c \triangleq \mathbf{H}_k^c(\overline{\mathbf{T}}^c, \mathbf{R}_k)$, and $\eta$ is given by
\begin{equation}\label{eq:nonh_eta}
    \eta = \Big\lVert\sum_{j=1}^K \sum_{c=1}^C \sum_{c^\prime=1}^C (\overline{\mathbf{H}}_j^c)^\hermconj\overline{\mathbf{\Phi}}_j\left(\mathbf{I}+\overline{\mathbf{\Gamma}}_j\right)\overline{\mathbf{\Phi}}_j^\hermconj\overline{\mathbf{H}}_j^{c^\prime}\Big\rVert_{\rm F}.
\end{equation}
To enforce power constraints on $\mathbf{Q}_k^c$, we first compute the transmission power with beamforming matrix $\mathbf{Q}_k^c$ as
\begin{equation}\label{eq:trQxQ}
    P_Q = \sum\nolimits_{k=1}^K\sum\nolimits_{c=1}^C \trace\left(\left(\mathbf{Q}_k^c\right)^\hermconj \mathbf{Q}_k^c\right).
\end{equation}
Scaling $\mathbf{Q}_k^c$ using $P_Q$, we obtain the expression of the constrained beamforming matrix $\mathbf{W}_k^c$, given by
\begin{equation}\label{eq:W}
    \mathbf{W}_k^c = \mathbf{Q}_k^c\min\left\{\sqrt{P_{\max} / P_Q}, 1\right\}.
\end{equation}

Given the closed-form expression of $\mathbf{W}_k^c$ in~\eqref{eq:W}, we show how to compute $\underline{\mathbf{W}}$ distributively under the DBP architecture in the sequel. First, we compute $\eta$ in a distributed manner. Since $\mathbf{I} + \overline{\mathbf{\Gamma}}_k \succ \mathbf{0}$, we perform the eigenvalue decomposition (EVD) to obtain $\mathbf{\Lambda}_k \in \mathbb{R}_{+}^{d \times d}$, a diagonal matrix of eigenvalues, and $\mathbf{\Xi}_k \in \mathbb{C}^{d \times d}$, whose columns are the corresponding eigenvectors. The CU broadcasts $\mathbf{\Lambda}_k$ and $\mathbf{\Xi}_k$ to all DUs, and each DU computes $\mathbf{P}_k^c$ as
\begin{equation}
    \mathbf{P}_k^c \triangleq (\overline{\mathbf{H}}_k^c)^\hermconj \overline{\mathbf{\Phi}}_k \mathbf{\Xi}_k \sqrt{\mathbf{\Lambda}_k}.
\end{equation}
Then, the value $\tilde{\mathbf{P}}_{kj} \triangleq \mathsf{Mul}(\mathbf{P}_k^c, \mathbf{P}_j^c)$ similar to the process in Fig.~\ref{fig:mul}, and the CU computes $\eta$ as
\begin{equation}\label{eq:eta_calc}
    \eta = \sqrt{\sum\nolimits_{j=1}^K \sum\nolimits_{k=1}^K \trace\left(\tilde{\mathbf{P}}_{kj}^\hermconj \tilde{\mathbf{P}}_{kj}\right)}.
\end{equation}
Note that~\eqref{eq:eta_calc} is equivalent to~\eqref{eq:nonh_eta} and is derived using the trace property of the Frobenius norm.

\begin{figure}[H]
    \vspace{-0.4cm}
    \centering
    \includegraphics[width=0.45\textwidth]{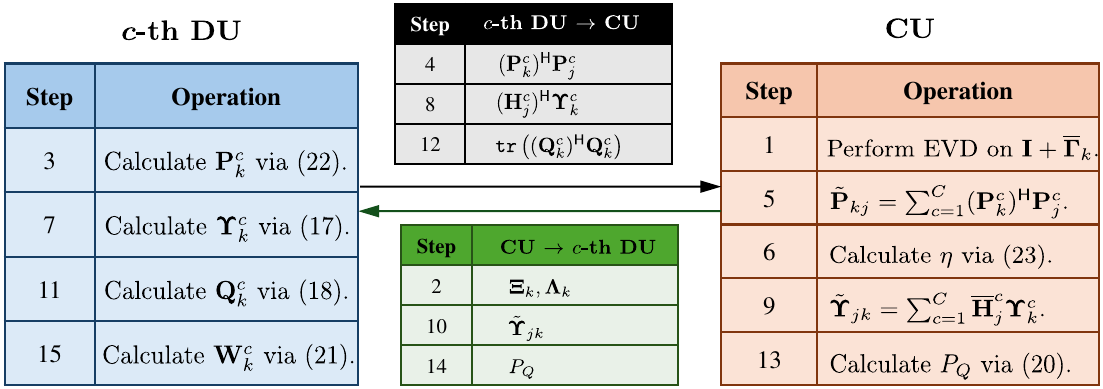}
    \caption{The decentralized calculation of $\underline{\mathbf{W}}$ under the DBP architecture.}
    \vspace{-0.1cm}
    \label{fig:bf}
\end{figure}

Next, we use the previously computed $\eta$ to calculate $\mathbf{W}_k^c$. First, we compute the extrapolated beamforming matrices $\mathbf{\Upsilon}_k^c$ at the $c$-th DU by~\eqref{eq:Upsilon}. Then, we compute $\tilde{\mathbf{\Upsilon}}_{jk}$ similar to Fig.~\ref{fig:mul} and compute $\mathbf{Q}_k^c$ via~\eqref{eq:extr_q}. To calculate the value of $P_Q$, each DU locally computes $\trace(\left(\mathbf{Q}_k^c\right)^\hermconj \mathbf{Q}_k^c)$ and transmits the result to the CU. The CU then aggregates the received values by~\eqref{eq:trQxQ}. Once $P_Q$ is obtained, it is broadcasted to all DUs, and $\mathbf{W}^c_k$ can be computed at the $c$-th DU by~\eqref{eq:W}. The aforementioned process is summarized in Fig.~\ref{fig:bf}, where only the $c$-th DU is shown for brevity.

\subsection{Update Step of $\underline{\mathbf{T}}$}\label{subsec:tx_pos}
In this step, we aim to optimize the transmit FA positions $\underline{\mathbf{T}}$ with the fixed $\underline{\mathbf{\Gamma}}$, $\underline{\mathbf{\Phi}}$, and $\underline{\mathbf{W}}$. The optimization problem~\eqref{opt:obj_quadratic} reduces to
\begin{equation}\label{opt_tx_position:iteration}          
    \underset{\underline{\mathbf{T}}}{\max}\ \ f_{\rm Quad}\left(\underline{\mathbf{T}}\right)\qquad{\text{s. t.}}\ \eqref{opt:tx_antenna_constraint}.
\end{equation} 
Unlike the update steps for $\underline{\mathbf{\Gamma}}$, $\underline{\mathbf{\Phi}}$, and $\underline{\mathbf{W}}$, the objective function $f_{\rm Quad}\left(\underline{\mathbf{T}}\right)$ remains non-concave w.r.t. $\underline{\mathbf{T}}$. To address this problem, we adopt the MM algorithm~\cite{sunMajorizationMinimizationAlgorithmsSignal2017} that iteratively finding a series of concave lower bounds for the non-concave function $f_{\rm Quad}\left(\underline{\mathbf{T}}\right)$, known as the \textit{surrogate function}. 

Specifically, each MM iteration consists of a \textit{majorization step}, followed by a \textit{maximization step}. In the \textit{majorization step}, the \textit{surrogate function} is computed using the second order Taylor expansion~\cite[Lemma~12]{sunMajorizationMinimizationAlgorithmsSignal2017}, given by
\begin{multline}\label{eq:opt_tx_position_surr}
    h^{\mathtt{Tx}}(\underline{\mathbf{T}}\big\vert\underline{\overline{\mathbf{T}}}) = -\frac{\delta^{\mathtt{Tx}}}2\sum\nolimits_{c=1}^C\vect(\mathbf{T}^c)^\transpose \vect(\mathbf{T}^c) \\
    + \sum_{c=1}^C\left(\nabla_{\vect(\mathbf{T}^c)} f_{\rm Quad}^\transpose(\overline{\mathbf{T}}^c) + \delta^{\mathtt{Tx}}\vect(\overline{\mathbf{T}}^c)^\transpose\right)\vect(\mathbf{T}^c),
\end{multline}
with $ h^{\mathtt{Tx}}(\underline{\mathbf{T}}\big\vert\underline{\overline{\mathbf{T}}}) \leq f_{\rm Quad}(\underline{\mathbf{T}})$. Here, $\delta^\mathtt{Tx}$ is selected such that $\delta^{\mathtt{Tx}} \geq \lambda_{\max}\big(\nabla_{\vect\left(\mathbf{T}\right)}^2 f_{\rm Quad}\left(\mathbf{\mathbf{T}}\right)\big)$ and it is calculated by $\delta^{\mathtt{Tx}} = \underset{1\leq c\leq C}{\max}\underset{1\leq m\leq M_c}{\max} \delta^{\mathtt{Tx}}_{m, c}$, where $\delta^{\mathtt{Tx}}_{m, c}$ is given by~\eqref{eq:opt_tx_position_delta} at the bottom of the page, 
\addtocounter{equation}{1}
and $\tilde{\mathbf{W}}_{kj} \triangleq \mathsf{Mul}(\overline{\mathbf{W}}_k^c, \overline{\mathbf{W}}_j^c)$. The entries of $\nabla_{\vect\left(\mathbf{T}^c\right)} f_{\rm Quad}\left(\underline{\mathbf{T}}\right)$ are given by
\begin{equation}\label{eq:opt_tx_position_dfdxyz}
    \frac{\partial f_{\rm Quad}}{\partial \mathbf{t}_m^c} = -\frac{4\pi}{\lambda} \sum_{k=1}^K \sum_{q=1}^{L_k^\mathtt{Tx}} \Big\lvert[\mathbf{D}_{k, c}^\mathtt{Tx}]_{mq}\Big\rvert \sin(\xi_{kmq, c}^{\mathtt{Tx}}) \mathbf{g}_{kq}^\mathtt{Tx},
\end{equation}
where $\mathbf{D}_{k, c}^\mathtt{Tx}$ and $\xi_{kmq, c}^{\mathtt{Tx}}$ are given by
\begin{multline}\label{eq:opt_tx_position_D}
    \mathbf{D}_{k, c}^\mathtt{Tx} \triangleq \Big(\frac{\partial f_{\rm Quad}}{\partial \mathbf{G}_k^c}\Big)^\transpose = \sqrt{\alpha_k}\overline{\mathbf{W}}_k^c\left(\mathbf{I} + \overline{\mathbf{\Gamma}}_k\right)\overline{\mathbf{F}}_k^\hermconj\mathbf{\Sigma}_k \\
    - \sum\nolimits_{j=1}^K\overline{\mathbf{W}}_j^c\tilde{\mathbf{G}}_{kj}^\hermconj\mathbf{\Sigma}_k^\hermconj\overline{\mathbf{F}}_k\overline{\mathbf{\Phi}}_k\left(\mathbf{I} + \overline{\mathbf{\Gamma}}_k\right)\overline{\mathbf{\Phi}}_k^\hermconj\overline{\mathbf{F}}_k^\hermconj\mathbf{\Sigma}_k
\end{multline}
and
\begin{align}\label{eq:opt_tx_position_xi}
    \xi_{kmq, c}^{\mathtt{Tx}} = \angle&[\mathbf{D}_{k, c}^\mathtt{Tx}]_{mq} + \frac{2\pi}{\lambda}\left(x_{m, c}^\mathtt{Tx}\cos\theta_{kq}^\mathtt{Tx}\cos\phi_{kq}^\mathtt{Tx} \right.\nonumber \\
    &\left.+ y_{m, c}^\mathtt{Tx}\cos\theta_{kq}^\mathtt{Tx}\sin\phi_{kq}^\mathtt{Tx} + z_{m, c}^\mathtt{Tx}\sin\theta_{kq}^\mathtt{Tx}\right),
\end{align}
respectively.

In the \textit{maximization step}, we determine the optimal $\underline{\mathbf{T}}$ by solving the following problem:
\begin{equation}\label{opt_tx_position:problem_mm}
    \underset{\underline{\mathbf{T}}}{\max}\ \ h^{\mathtt{Tx}}\left(\underline{\mathbf{T}} \big\vert \overline{\underline{\mathbf{T}}}\right),\qquad {\text{s. t.}}\ \eqref{opt:tx_antenna_constraint}.
\end{equation}
By observing that the problem~\eqref{opt_tx_position:problem_mm} has only cuboid boundaries, the closed-form solution can be obtained by projecting the unconstrained optimum $\mathbf{T}^\star$ onto the cuboid region~\cite{maMIMOCapacityCharacterization2024}. The unconstrained optimum $\mathbf{T}^\star$ is given by
\begin{equation}\label{eq:opt_tx_position_closed_form}
    \mathbf{T}^{c, \star} = \overline{\mathbf{T}}^c + \frac{1}{\delta^{\mathtt{Tx}}} \nabla_{\mathbf{T}^c} f_{\rm Quad}(\overline{\mathbf{T}}^c).
\end{equation}
Projecting $\mathbf{T}^\star$ onto the cuboid region yields the closed-form optimal solution, given by
\begin{equation}\label{eq:opt_tx_position_proj}
	p_{m, c}^\mathtt{Tx} = \min\Big(\max\left(p_{m, c}^{\mathtt{Tx}, \star}, p_{m, c}^{\min}\right), p_{m, c}^{\max}\Big).
\end{equation}
Here, $p$ denotes a spatial coordinate, which can be $x$, $y$, or $z$. The terms $p_{m, c}^{\min}$ and $p_{m, c}^{\max}$ denote the lower and upper boundaries of the cuboid movable region for the $m$-th transmit FA at the $c$-th DU, respectively. The process described above is an iteration of the proposed decentralized MM algorithm for optimizing $\underline{\mathbf{T}}$, and it is executed until convergence.

\begin{figure}[h]
    \centering
    \vspace{-0.4cm}
    \includegraphics[width=0.45\textwidth]{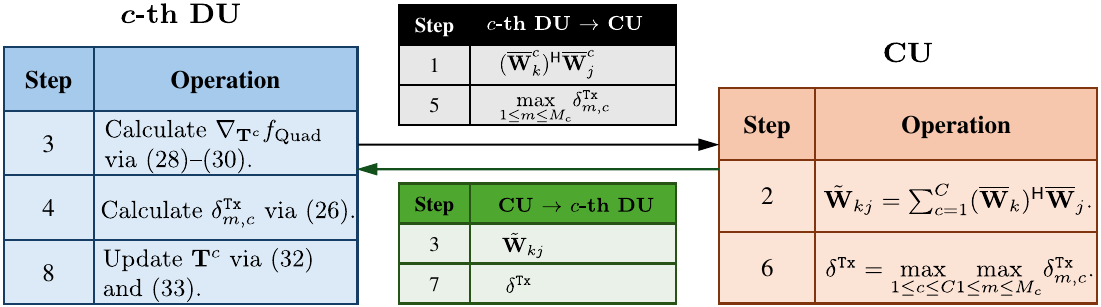}
    \caption{The decentralized update of $\underline{\mathbf{T}}$ under the DBP architecture.}
    \vspace{-0.3cm}
    \label{fig:tx_pos}
\end{figure}
Next, we show the update of $\underline{\mathbf{T}}$ under the DBP architecture. First, we compute the coefficients of the \textit{surrogate function} $h^{\mathtt{Tx}}(\underline{\mathbf{T}}\big\vert\underline{\overline{\mathbf{T}}})$ given by~\eqref{eq:opt_tx_position_surr}. This begins with the distributed computation of $\tilde{\mathbf{W}}_{kj}$ as illustrated by Fig.~\ref{fig:mul}. After the CU broadcasts $\tilde{\mathbf{W}}_{kj}$ to all DUs, the entries of $\nabla_{\mathbf{T}^c}f_{\rm Quad}(\overline{\mathbf{T}}^c)$ are calculated by~\eqref{eq:opt_tx_position_dfdxyz},~\eqref{eq:opt_tx_position_D}, and~\eqref{eq:opt_tx_position_xi} at each DU. Next, to compute $\delta^\mathtt{Tx}$, each DU evaluates $\delta^{\mathtt{Tx}}_{m, c}$ by~\eqref{eq:opt_tx_position_delta} and selects the greatest value to send to the CU, which determines the largest and sets it as $\delta^{\mathtt{Tx}}$. With these coefficients computed, the optimal solution for $\underline{\mathbf{T}}$ is computed at the DUs using~\eqref{eq:opt_tx_position_closed_form} and~\eqref{eq:opt_tx_position_proj}. The aforementioned process is summarized in Fig.\,\ref{fig:tx_pos}, where only the $c$-th DU is shown for brevity.

\begin{figure*}[hb]
    \normalsize
    \setcounter{MYtempeqncnt}{\value{equation}}
    \setcounter{equation}{34}
    \vspace{-1\baselineskip}
    \hrulefill
    \begin{equation}\label{eq:opt_rx_position_delta}
        \begin{aligned}
            \delta^{\mathtt{Rx}}_k = \underset{1\leq n\leq N}{\max}\frac{24\pi^2}{\lambda^2} &L_k^r\left(\left(\sum\nolimits_{j=1}^N\Big\lvert[\overline{\mathbf{\Phi}}_k]_n\left(\mathbf{I} + \overline{\mathbf{\Gamma}}_k\right)[\overline{\mathbf{\Phi}}_k]_j^\hermconj\Big\rvert + \sqrt{N}\Big\lVert[\overline{\mathbf{\Phi}}_k]_n\left(\mathbf{I} + \overline{\mathbf{\Gamma}}_k\right)\overline{\mathbf{\Phi}}_k^\hermconj\Big\rVert_2\right)\right.\\
            &\left.\times \Big\lVert\sum\nolimits_{t=1}^K \mathbf{\Sigma}_k \tilde{\mathbf{G}}_{kt}\tilde{\mathbf{G}}_{kt}^\hermconj \mathbf{\Sigma}_k^\hermconj\Big\rVert_2 + \sqrt{\frac{\alpha_k}{L_k^r}}\Big\lVert[\mathbf{\Phi}_k]_n\left(\mathbf{I} + \overline{\mathbf{\Gamma}}_k\right)\tilde{\mathbf{G}}_{kk}^\hermconj\mathbf{\Sigma}_k^\hermconj\Big\rVert_2\right).
        \end{aligned}    
    \end{equation}
    \setcounter{equation}{\value{MYtempeqncnt}}
\end{figure*}

\subsection{Update Step of $\underline{\mathbf{R}}$}\label{subsec:rx_pos}
In this step, our target is to optimize the positions of the receive FAs $\underline{\mathbf{R}}$ with fixed $\underline{\mathbf{T}}$, $\underline{\mathbf{W}}$, $\underline{\mathbf{\Gamma}}$, and $\underline{\mathbf{\Phi}}$. The objective function $f_{\rm Quad}\left(\underline{\mathbf{R}}\right)$ is reformulated as
\begin{equation}\label{eq:opt_rx_position_sum}
    f_{\rm Quad}\left(\underline{\mathbf{R}}\right) = \sum\nolimits_{k=1}^K f_{\rm Quad}\left(\mathbf{R}_k\right).
\end{equation}
Since the terms of the right hand side of~\eqref{eq:opt_rx_position_sum} do not couple with each other, it is feasible to optimize $f_{\rm Quad}\left(\mathbf{R}_k\right)$ independently. Therefore, we simply provide the update step of $\mathbf{R}_k$ in the remainder of this subsection.

Similar to the update step of $\underline{\mathbf{T}}$, we use the MM algorithm to optimize $\mathbf{R}_k$. In the \textit{majorization step}, we construct the \textit{surrogate function} as
\begin{multline}\label{eq:opt_rx_position_surr}
    h^{\mathtt{Rx}}_k\left(\mathbf{R}_k\big\vert\overline{\mathbf{R}}_k\right) = -\frac{\delta^{\mathtt{Rx}}_k}2\vect\left(\mathbf{R}_k\right)^\transpose \vect\left(\mathbf{R}_k\right) \\
    + \left(\nabla_{\vect\left(\mathbf{R}_k\right)} f_{\rm Quad}^\transpose \left(\overline{\mathbf{R}}_k\right)+ \delta^{\mathtt{Rx}}_k\vect\left(\overline{\mathbf{R}}_k\right)^\transpose\right)\vect\left(\mathbf{R}_k\right).
\end{multline}
Here, the expression of $\delta^\mathtt{Rx}_k$ is given by~\eqref{eq:opt_rx_position_delta} at the bottom of the page, and the entries of $\nabla_{\vect\left(\mathbf{R}_k\right)} f_{\rm Quad}\left(\mathbf{R}_k\right)$ are given by
\addtocounter{equation}{1}
% \begin{subequations}\label{eq:opt_rx_position_dfdxyz}
%     \begin{align}
%         \frac{\partial f_{\rm Quad}}{\partial x_{kn}^\mathtt{Rx}} &= -\frac{4\pi}{\lambda}\sum_{k=1}^K\sum_{q=1}^{L_k^\mathtt{Rx}}\Big\lvert[\mathbf{D}_k^\mathtt{Rx}]_{nq}\Big\rvert\cos\theta_{kq}^\mathtt{Rx}\cos\phi_{kq}^\mathtt{Rx}\sin\left(\xi_{knq}^{\mathtt{Rx}}\right), \\
%         \frac{\partial f_{\rm Quad}}{\partial y_{kn}^\mathtt{Rx}} &= -\frac{4\pi}{\lambda}\sum_{k=1}^K\sum_{q=1}^{L_k^\mathtt{Rx}}\Big\lvert[\mathbf{D}_k^\mathtt{Rx}]_{nq}\Big\rvert\cos\theta_{kq}^\mathtt{Rx}\sin\phi_{kq}^\mathtt{Rx}\sin\left(\xi_{knq}^{\mathtt{Rx}}\right), \\
%         \frac{\partial f_{\rm Quad}}{\partial z_{kn}^\mathtt{Rx}} &= -\frac{4\pi}{\lambda}\sum_{k=1}^K\sum_{q=1}^{L_k^\mathtt{Rx}}\Big\lvert[\mathbf{D}_k^\mathtt{Rx}]_{nq}\Big\rvert\sin\theta_{kq}^\mathtt{Rx}\sin\left(\xi_{knq}^{\mathtt{Rx}}\right),
%     \end{align}
% \end{subequations}
\begin{equation}\label{eq:opt_rx_position_dfdxyz}
    \frac{\partial f_{\rm Quad}}{\partial \mathbf{r}_{kn}} = \frac{4\pi}{\lambda}\sum\nolimits_{k=1}^K\sum\nolimits_{q=1}^{L_k^\mathtt{Rx}}\Big\lvert[\mathbf{D}_k^\mathtt{Rx}]_{nq}\Big\rvert \sin(\xi_{knq}^{\mathtt{Rx}}) \mathbf{f}_{kq}^\mathtt{Rx}.
\end{equation}
The expressions of $\mathbf{D}_k^\mathtt{Rx}$ and $\xi_{knq}^{\mathtt{Rx}}$ are given by
\begin{align}\label{eq:opt_rx_position_D}
    \mathbf{D}^{\mathtt{Rx}}_k \triangleq& \Big(\frac{\partial f_{\rm Quad}}{\partial \mathbf{F}_k}\Big)^\transpose = \sqrt{\alpha_k}\overline{\mathbf{\Phi}}_k\left(\mathbf{I} + \overline{\mathbf{\Gamma}}_k\right)\tilde{\mathbf{G}}_{kk}^\hermconj \mathbf{\Sigma}_k^\hermconj \nonumber\\
    &- \overline{\mathbf{\Phi}}_k \left(\mathbf{I} + \overline{\mathbf{\Gamma}}_k\right) \overline{\mathbf{\Phi}}_k^\hermconj \overline{\mathbf{F}}_k^\hermconj\mathbf{\Sigma}_k \sum\nolimits_{j=1}^K \tilde{\mathbf{G}}_{kj}\tilde{\mathbf{G}}_{kj}^\hermconj \mathbf{\Sigma}_k^\hermconj.
\end{align}
and
\begin{align}
    \xi_{knq}^{\mathtt{Rx}} =& \angle[\mathbf{D}^{\mathtt{Rx}}_k]_{nq} + \frac{2\pi}{\lambda}\left(x_{kn}^\mathtt{Rx}\cos\theta_{kq}^\mathtt{Rx}\cos\phi_{kq}^\mathtt{Rx} \right.\nonumber \\
    &\left.+ y_{kn}^\mathtt{Rx}\cos\theta_{kq}^\mathtt{Rx}\sin\phi_{kq}^\mathtt{Rx} + z_{kn}^\mathtt{Rx}\sin\theta_{kq}^\mathtt{Rx}\right),
\end{align}
respectively. In the \textit{maximization step}, we calculate the optimal $\mathbf{R}_k$ by solving the optimization problem given by
\begin{equation}\label{opt_rx_position:problem_mm}
    \underset{\mathbf{R}_k}{\max}\ \ h_k^{\mathtt{Rx}}(\mathbf{R}_k \big\vert \overline{\mathbf{R}}_k),\qquad {\text{s. t.}}\ \eqref{opt:rx_antenna_constraint}.
\end{equation}
We first derive the solution to the problem~\eqref{opt_rx_position:problem_mm} without the constraint~\eqref{opt:rx_antenna_constraint}, given by
\begin{equation}\label{eq:opt_rx_position_closed_form}
    \mathbf{R}_k^\star = \overline{\mathbf{R}}_k + \frac{1}{\delta^{\mathtt{Rx}}_k} \nabla_{\mathbf{R}_k} f_{\rm Quad}\left(\overline{\mathbf{R}}_k\right).
\end{equation}
Then, the closed form solution to the problem~\eqref{opt_rx_position:problem_mm} can be obtained by projecting $\mathbf{R}_k^\star$ onto the cuboid regions, given by
\begin{equation}\label{eq:opt_rx_position_proj}
	p_{kn}^\mathtt{Rx} = \min\Big(\max\left(p_{kn}^{\mathtt{Rx}, \star}, p_{kn}^{\min}\right), p_{kn}^{\max}\Big).
\end{equation}

The calculation of $\mathbf{R}_{kn}$ can be directly executed at the CU. First, we compute the coefficients of the \textit{surrogate function} $h^{\mathtt{Rx}}_k(\mathbf{R}_k\big\vert\overline{\mathbf{R}}_k)$. Leveraging the $\tilde{\mathbf{G}}_{kj}$ computed previously, the CU computes the entries of $\nabla_{\mathbf{R}_k} f_{\rm Quad}(\overline{\mathbf{R}}_k)$ and $\delta_k^\mathtt{Rx}$ using~\eqref{eq:opt_rx_position_dfdxyz} and~\eqref{eq:opt_rx_position_delta}, respectively. Then, the optimal solution to problem~\eqref{opt_rx_position:problem_mm} is computed at the CU via~\eqref{eq:opt_rx_position_closed_form} and~\eqref{eq:opt_rx_position_proj}. The key steps of the above decentralized DBP-based algorithm are summarized in Algorithm~\ref{alg:dec_opt_overall}.
\vspace{-0.2cm}
\begin{algorithm}
    \caption{Decentralized DBP-based Algorithm}\label{alg:dec_opt_overall}
    \begin{algorithmic}[1]
        \Require {\small $C$, $M$, $N$, $K$, $P_{\max}$, $\alpha_k$, $\mathbf{\Sigma}_k$, $L_k^\mathtt{Tx}$, $L_k^\mathtt{Rx}$, $\theta_{ki}^\mathtt{Tx}$, $\phi_{ki}^\mathtt{Tx}$, $\theta_{kj}^\mathtt{Rx}$, $\phi_{kj}^\mathtt{Rx}$.}
        \State Initialize $\underline{\mathbf{W}}$, $\underline{\mathbf{T}}$, and $\underline{\mathbf{R}}$ to corresponding feasible values.
        \Repeat
        \State Update $\underline{\mathbf{\Phi}}_k$ and $\underline{\mathbf{\Gamma}}_k$ according to Section~\ref{subsec:auxiliary}.
        \State Update $\underline{\mathbf{W}}$ according to Section~\ref{subsec:W}.
        \State Update $\underline{\mathbf{T}}$ according to Section~\ref{subsec:tx_pos}.
        \State Update $\underline{\mathbf{R}}$ according to Section~\ref{subsec:rx_pos}.
        \Until{the value of $R$ converges.}
        \Ensure $\underline{\mathbf{W}}$, $\underline{\mathbf{T}}$, $\underline{\mathbf{R}}$.
    \end{algorithmic}
\end{algorithm}
\vspace{-0.3cm}

\textit{Remark:} In the proposed decentralized BCA-based algorithm, the dimension of any data transmitted between the CU and DUs is independent of $M$, ensuring low \textit{communication cost}. Moreover, the dimension of all stored matrices is also independent of $M$, which guarantees low \textit{storage cost}.

\section{Simulation}\label{sec:sim}
\begin{table}[tb]
	\centering
	\caption{Key Simulation Parameters}
	\begin{small}
	\begin{tabular}{|l|l|}
		\hline
		\textbf{Parameter} & \textbf{Value} \\ \hline
		Number of channel realizations & $S = 200$ \\
		Number users &  $K = 6$ \\
		User priority & $\alpha_k = 1$ \\
		Number of FAs at each user &  $N = 4$     \\
		Number of parallel data streams &  $d = 4$      \\
		Carrier freqency & $f_c = 28$ GHz \\ 
		% Carrier wavelength & $\lambda = 10.7$ mm \\ 
		Transmit power budget & $P_{\max} = 20$ dBm \\
		Noise power & $\sigma_k^2 = -80$ dBm \\
		Miminum user distance from the BS & $d_{\min} = 20$ m \\
		Maximum user distance from the BS & $d_{\max} = 100$ m \\
		Distance from the BS to user $k$ & $d_k^2 \sim \mathcal{U}[d_{\min}^2, d_{\max}^2]$ \\
		% Reference distance & $d_0 = 1$ m \\
		Pathloss exponent & $\varrho = 3.67$ \\
		Pathloss at reference distance $d_0 = 1$ m & $T_0 = -61.4$ dB \\
		Elevation/Azimuth AoD & $\theta_{kq}^\mathtt{Tx}, \phi_{kq}^\mathtt{Tx} \sim \mathcal{U}[0, \pi)$ \\
		Elevation/Azimuth AoA & $\theta_{kq}^\mathtt{Rx}, \phi_{kq}^\mathtt{Rx} \sim \mathcal{U}[0, \pi)$ \\
		Number of transmit/receive paths & $L_k^\mathtt{Tx} = L_k^\mathtt{Rx} = 3$ \\
		\hline
	\end{tabular}
	\end{small}%
    \vspace{-1\baselineskip}
	\label{tab:sim_param}
\end{table}

In this section, we evaluate the performance of the FA-assisted MU-MIMO systems with DBP architecture using the proposed decentralized BCA-based algorithm. Unless otherwise specified, the key simulation parameters follow the settings in Table~\ref{tab:sim_param}. Specifically, the pathloss of user $k$ is calculated as $\kappa(d_k) = T_0 (d_k / d_0)^{-\varrho}$, and the PRM is diagonal with entries following $[\mathbf{\Sigma}_k]_{qq} \sim \mathcal{CN}(0, \kappa(d_k)/L)$. We denote the system described in Section~\ref{sec:system} as transmit and receive FA (TRFA), and refer to the system with fixed $\lambda/2$-spaced antennas as FPA. Moreover, ``C'' denotes the BCA-based centralized algorithm~\cite{liao2025joint}, and ``D'' represents the proposed decentralized algorithm using the DBP architecture.

\begin{figure}[tb]
    \centering
    \includegraphics[width=0.47\textwidth]{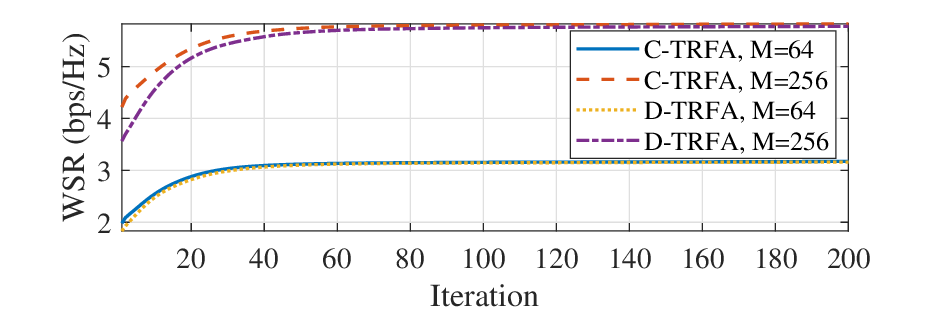}
    \caption{Convergence behaviors of the proposed algorithms.}
    \vspace{-0.3cm}
    \label{fig:conv}
\end{figure}

\begin{table}[tb]
    \centering
    \caption{WSR Performance Comparison}
    \begin{small}
    \begin{tabular}{|c|c|cccc|}
        \hline
        \multirow{2}{*}{$M$} & \multirow{2}{*}{\begin{tabular}[c]{@{}c@{}}$P_{\max}$ \\ (dBm)\end{tabular}} & \multicolumn{4}{c|}{WSR (bps/Hz)} \\ \cline{3-6} 
                                &      & \multicolumn{1}{c|}{C-FPA}  & \multicolumn{1}{c|}{D-FPA}  & \multicolumn{1}{c|}{C-TRFA} & \textbf{D-TRFA} \\ \hline
        \multirow{2}{*}{$64$}  & $20$ & \multicolumn{1}{c|}{$2.27$} & \multicolumn{1}{c|}{$2.21$} & \multicolumn{1}{c|}{$3.23$} & $\mathbf{3.22}$ \\ \cline{2-6} 
                                & $30$ & \multicolumn{1}{c|}{$7.30$} & \multicolumn{1}{c|}{$7.19$} & \multicolumn{1}{c|}{$8.75$} & $\mathbf{8.66}$ \\ \hline
        \multirow{2}{*}{$256$} & $20$ & \multicolumn{1}{c|}{$4.68$} & \multicolumn{1}{c|}{$4.70$} & \multicolumn{1}{c|}{$5.95$} & $\mathbf{5.91}$ \\ \cline{2-6} 
                                & $30$ & \multicolumn{1}{c|}{$13.8$} & \multicolumn{1}{c|}{$13.5$} & \multicolumn{1}{c|}{$15.3$} & $\mathbf{15.1}$ \\ \hline
        \end{tabular}
    \end{small}%
    \vspace{-0.8\baselineskip}
    \label{tab:perf}
\end{table}

\begin{table}[tb]
    \caption{CPU Time Saved Compared with Centralized \\ Implementation (\%)}%\vspace{0.2cm}
    \centering
    \begin{small}
        \begin{tabular}{|c|cc|cc|}
            \hline
            $M$  & \multicolumn{2}{c|}{$64$}                & \multicolumn{2}{c|}{$256$}               \\ \hline
            $C$  & \multicolumn{1}{c|}{$4$}      & $16$     & \multicolumn{1}{c|}{$4$}      & $16$     \\ \hline
            FPA  & \multicolumn{1}{c|}{$89.6\%$} & $98.3\%$ & \multicolumn{1}{c|}{$99.2\%$} & $99.5\%$ \\ \hline
            TRFA & \multicolumn{1}{c|}{$69.4\%$} & $87.9\%$ & \multicolumn{1}{c|}{$71.9\%$} & $82.4\%$ \\ \hline
        \end{tabular}
    \end{small}%
    \vspace{-0.7cm}
    \label{tab:time_dr}
\end{table}

The convergence behaviors are shown in Fig.\,\ref{fig:conv}, where all curves increase with the number of iterations and eventually stabilize, confirming the convergence of the proposed algorithm. Furthermore, both algorithms require the similar number of iterations to converge, demonstrating the efficiency of the decentralized approach.

As shown in Table~\ref{tab:perf}, we compare the WSR performance of the decentralized BCA-based algorithm with that of the centralized algorithm under various setups. The WSR loss of the decentralized algorithm is consistently negligible compared with its centralized counterpart. 
We illustrate the computational efficiency of the proposed algorithm by measuring central processing unit (CPU) time. Specifically, the total CPU time is recorded for the centralized algorithm, while for the decentralized algorithm under the DBP architecture, the CPU time is computed as the sum of the CU's CPU time and the maximum running time among all DUs. As shown in Table~\ref{tab:time_dr}, the DBP architecture significantly reduces computation time by at least $69.4\%$ compared with the centralized algorithm across all system configurations. This improvement stems from the parallel processing capability of the DBP architecture, where each DU independently solves a smaller-scale problem. For a fixed number of transmit FAs $M$, increasing the number of DUs $C$ enhances parallelism, thereby saving more computational time.

\section{Conclusion and Future Work}\label{sec:conclusion}
To mitigate the high complexity introduced by the FAS, we incorporated the DBP architecture into FA-assisted MU-MIMO systems, and investigated the joint optimization of beamforming and FA positions to maximize WSR. We proposed a novel decentralized BCA-based algorithm with a decentralized inverse-free algorithm for beamforming and a distributed MM algorithm for FA position optimization to solve the problem. Simulation results demonstrated that the proposed algorithm significantly reduced computational costs with negligible WSR loss compared to centralized counterpart. Future directions include the distributed channel estimation and signal detection in FA-assisted MIMO systems with DBP.

\bibliographystyle{IEEEtran}

\begin{spacing}{0.85}
\bibliography{IEEEabrv, reference}
\end{spacing}

\end{document}